\documentclass[nofootinbib, showpacs]{revtex4}

\usepackage[utf8]{inputenc}
\usepackage[T1]{fontenc}
\usepackage{amsmath,amssymb}

\DeclareMathOperator{\sgn}{sgn}

\begin{document}
\title{Hybrid quantization of an inflationary model: The flat case}
\date{\today}

\author{Mikel Fern\'andez-M\'endez}
\email{m.fernandez.m@csic.es}
\affiliation{Instituto de Estructura de la Materia, IEM-CSIC, Serrano 121, 28006 Madrid, Spain}
\author{Guillermo A. Mena Marug\'an}
\email{mena@iem.cfmac.csic.es}
\affiliation{Instituto de Estructura de la Materia, IEM-CSIC, Serrano 121, 28006 Madrid, Spain}
\author{Javier Olmedo}
\email{jolmedo@fisica.edu.uy}
\affiliation{Instituto de F\'{i}sica, Facultad de Ciencias, Igu\'a 4225, esq.\ Mataojo, Montevideo, Uruguay}

\begin{abstract}
We present a complete quantization of an approximately homogeneous and isotropic universe with small scalar perturbations. We consider the case in which the matter content is a minimally coupled scalar field and the spatial sections are flat and compact, with the topology of a three-torus. The quantization is carried out along the lines that were put forward by the authors in a previous work for spherical topology. The action of the system is truncated at second order in perturbations. The local gauge freedom is fixed at the classical level, although different gauges are discussed and shown to lead to equivalent conclusions. Moreover, descriptions in terms of gauge-invariant quantities are considered. The reduced system is proven to admit a symplectic structure, and its dynamical evolution is dictated by a Hamiltonian constraint. Then, the background geometry is polymerically quantized, while a Fock representation is adopted for the inhomogeneities. The latter is selected by uniqueness criteria adapted from quantum field theory in curved spacetimes, which determine a specific scaling of the perturbations. In our hybrid quantization, we promote the Hamiltonian constraint to an operator on the kinematical Hilbert space. If the zero mode of the scalar field is interpreted as a relational time, a suitable ansatz for the dependence of the physical states on the polymeric degrees of freedom leads to a quantum wave equation for the evolution of the perturbations. Alternatively, the solutions to the quantum constraint can be characterized by their initial data on the minimum-volume section of each superselection sector. The physical implications of this model will be addressed in a future work, in order to check whether they are compatible with observations.
\end{abstract}

\pacs{04.60.Pp, 04.60.Kz, 98.80.Qc }

\maketitle

\section{Introduction}

In the standard analysis of cosmological inflation and primordial fluctuations, one combines the classical description of a homogeneous and isotropic background universe with the quantum treatment of perturbations---the inhomogeneities---propagating on it~\cite{Mukhanov05,inflation}. Although extremely successful, this procedure is not completely satisfactory, in part because of the theoretical tension inherent to any semiclassical model. Moreover, it has been argued that the magnification of scales during inflation could amplify the effects of high-energy physics~\cite{MB01}, not to mention that the cosmological singularities of the classical theory persist in the semiclassical approach. However, the attempts to overcome these problems are hindered by the absence of a complete quantum theory of gravity. The promising development of Loop Quantum Gravity (LQG)~\cite{LQG}, a nonperturbative, background-independent, canonical quantization of General Relativity, may indicate some progress in that direction. The application of LQG techniques to symmetry-reduced systems, known as Loop Quantum Cosmology (LQC)~\cite{LQC}, has yielded noteworthy results as well. First of all, it has succeeded in providing a consistent quantization of the Friedmann-Lema\^\i tre-Robertson-Walker (FLRW) model, in which the big-bang singularity is replaced with a big bounce~\cite{APS06b,APS06c,moreFLRW,MOP11}. Remarkably, the quantum dynamics of this system can be approximated very well by some effective classical dynamics in physically meaningful situations~\cite{effective}. More general models have also been successfully quantized, such as anisotropic~\cite{anisotropic} and inhomogeneous ones~\cite{hybrid1,hybrid2}. However, only in the homogeneous and isotropic case has the effective dynamics been derived analytically and tested thoroughly.

In LQC, the effective dynamics of an FLRW universe filled with a minimally coupled, massive scalar field drives the system to a phase of slow-roll inflation with a high probability---for natural values of the field mass---\cite{AS10}, in contrast to what is expected in General Relativity. The active study of inflation and cosmological perturbations in the framework of LQC is partly motivated by the desire to search for testable predictions of the theory~\cite{predictions}. Some of the works on this issue introduce inverse-volume corrections~\cite{predictions, inverse}, holonomy corrections~\cite{holonomy}, or both~\cite{both} (always starting with certain constructions for the possible quantum modifications), in order to obtain effective constraints and equations that are argued to capture the dynamics of the full quantum theory in a good approximation. Our strategy is different: we truncate the classical action of the theory at second order in perturbations and then proceed to a full quantization of both the perturbations and the background~\cite{FMO12}.

This scheme is closer to that adopted by Agullo \textit{et al.}~\cite{AAN1,AAN2,AAN3} in a series of papers appeared soon after the publication of Ref.~\cite{FMO12}. Adhering to the use of the homogeneous part of the scalar field as a relational time, those authors have found that the quantum perturbations can be interpreted as fields propagating in a \emph{dressed} effective spacetime (see also Ref.~\cite{QFTonQST}). Note, however, that the truncation scheme adopted in those works is different from ours. As a consequence of the different approach, and in contrast with the results of those works, we succeed in obtaining a symplectic structure for the whole perturbed system (both before and after gauge fixing), as well as a Hamiltonian constraint that generates the reduced dynamics in the truncated model. The incorporation of second-order corrections to the homogeneous variables allows us to maintain a (constrained) Hamiltonian evolution even after reduction. Besides, the quantization attained in this manner does not rest on the introduction of any specific relational time: it is not necessary to deparametrize the system in order to reach a consistent dynamical description. Moreover, our approach provides a specific scaling of the perturbations in terms of the scale factor of the FLRW geometry. This scaling is essential in order to reach a privileged quantization with unitary semiclassical dynamics. Remarkably, such a scaling has not been adopted in Ref.~\cite{AAN1,AAN2,AAN3}. We are concerned that the lack of unitarity in the quantum dynamics of the perturbations, in the regime in which the standard description of quantum field theory in curved spacetimes should apply, could compromise further results if this scaling is not performed \emph{before the quantization}.

With this motivation, in this work we will discuss in detail the implementation of the hybrid approach~\cite{hybrid1,hybrid2} in the quantization of an FLRW model minimally coupled to a massive scalar field with scalar perturbations on flat (but compact) spatial slices. The case of spatial sections with positive curvature was already addressed in Ref.~\cite{FMO12}. We prove here that there is no obstacle in applying the same techniques to the flat case. In fact, this latter case is expectably simpler, since spatial curvature effects are not present. This also explains why, in a certain sense, it seemed natural to discuss first the case of spherical topology, as we did in Ref.~\cite{FMO12}, and then pass to the flat model. This flat scenario is specially important, because the cosmological measurements indicate that the universe is approximately flat~\cite{wmaplanck} (although this might be a mere consequence of inflation). Besides, we consider only scalar perturbations precisely because of their observational relevance. The decoupling from other kinds of perturbations makes this treatment consistent. On the other hand, the theoretical analysis of scalar perturbations is the most intricate one, inasmuch as vector perturbations are pure gauge, and tensor ones are gauge invariant and have simpler dynamical equations.

In the study of scalar cosmological perturbations, the consideration of gauge-invariant quantities avoids the dependence of the results on the identification of the background spacetime~\cite{Bardeen80} (see also Ref.~\cite{Langlois94} for a Hamiltonian treatment). One of those quantities is the Mukhanov-Sasaki (MS) variable~\cite{Mukhanov05,MSvariable}, especially useful in the case of a flat universe, because its evolution is then given by a Klein-Gordon (KG) equation with a background-dependent mass. This variable is usually taken as the starting point for quantization in the classical background provided by the FLRW spacetime. The corresponding Fock representation is determined by choosing the Bunch-Davies vacuum [which possesses the O(1,4) invariance of de Sitter spacetime]. In a genuine quantum formalism, this analysis has to be extended to incorporate appropriately the quantum nature of the background geometry.

A possible approach is provided by the hybrid quantization scheme. Hybrid quantization still represents the inhomogeneities \`a la Fock, whereas it adopts a polymeric representation for the background. In this way, one can define rigorously a kinematical Hilbert space in which the Hamiltonian constraint can be represented. This strategy was followed for the first time~\cite{hybrid1} in the Gowdy models~\cite{Gowdy74}, which include gravitational waves but retain the symmetry corresponding to two spatial Killing vectors. With a suitable parametrization, the Hamiltonian of these systems is a quadratic function of the field. Thus, Fock quantization is well suited to deal with it. No truncation is needed to arrive at linear dynamical equations for the inhomogeneities. Therefore, no perturbative truncation is needed: the treatment of the system is exact, providing the best arena to test the quantization methods and discuss their physical consequences. On the other hand, the complete loop quantization of the model would be an extremely ambitious task. Hybrid quantization is a compromise that allows to investigate the effects of discrete geometry in the homogeneous sector, while maintaining the infinite degrees of freedom. The analytic and numerical studies of the effective dynamics of the resulting theory showed that the big bang is replaced with a bounce in which the inhomogeneities can be amplified~\cite{hybrid2}.

It is known that the Fock quantization of a field theory in a curved spacetime~\cite{QFTCS} is plagued with ambiguities. Different representations of the same algebra are in general unitarily inequivalent. Nevertheless, it has been shown that the requirements of a symmetric vacuum (i.e., a vacuum with the same spatial symmetries of the field equations) and unitarily implementable dynamics suffice to overcome this problem in the case of a KG field with a time-dependent quadratic potential in (e.g.)\ a ultrastatic spacetime, assuming compact spatial sections of dimension equal or less than three. For this kind of field theories, all the representations with the mentioned properties belong to the same unitary equivalence class~\cite{CMOV11}. Remarkably, these criteria remove as well the ambiguity in the choice of fundamental variables in the following sense: if, by means of a local time-dependent linear canonical transformation, one arrives at other canonical pair describing the field, the requirements cannot be satisfied~\cite{CMOV12} (except in some situations in one spatial dimension in which, nonetheless, the physics is not affected by the transformation). It is worth noticing that any scaling of the field by a background function can be viewed as part of a time-dependent linear canonical transformation. The uniqueness result about the choice of fundamental variables, therefore, tells us how to split the field in a purely fieldlike part and a background-dependent part. In addition, a particular type of \emph{nonlocal} time-dependent linear canonical transformations has also been studied, namely those which, apart from being compatible with the symmetries, preserve the form of the equations of motion~\cite{CFMM13}. These transformations admit necessarily a unitary implementation~\cite{CFMM13}.

Naturally, these results apply as well to field theories in cosmological spacetimes if they can be interpreted as describing the propagation of a scalar field with time-dependent mass in an auxiliary ultrastatic spacetime. In fact, they were first found~\cite{uniqueGowdy} precisely in the Gowdy models, whose symmetry reduction leads to what can be regarded as a KG field with a background-dependent potential propagating on circles or two-spheres. The results were latter extended to general manifolds. The specially relevant case in which the spatial manifold is a three-torus was discussed in detail in Ref.~\cite{CCMMV12}. Even in the presence of subdominant corrections to the field equations in the ultraviolet limit, the results have been proven to hold in the three-sphere~\cite{FMOV12}, adapting the arguments of Ref.~\cite{3-sphere} (actually, similar conclusions can be reached for the three-torus). The relation of this quantization and the Hadamard one is discussed in Refs.~\cite{CMOV12,CMMV13}.

The rest of this article is organized as follows. In Sec.~\ref{sec:classical}, we introduce and reformulate appropriately the classical model: we propose two alternate gauge fixings and perform a canonical transformation to the preferred field variables in each of these gauge choices. The relation with the MS variable is also discussed. Sec.~\ref{sec:quantization} is devoted to the hybrid quantization of the system. In Sec.~\ref{subsec:hom}, we review the quantization of the unperturbed model, while we address the Fock quantization of the inhomogeneities in Sec.~\ref{subsec:inhom}. Once the total kinematical Hilbert space is constructed, we propose a prescription to promote the Hamiltonian constraint to an operator. Its solutions are studied in Sec.~\ref{sec:phys}, either by using the zero mode of the field as a relational time (Sec.~\ref{subsec:time}) or in terms of the constant FLRW-volume sections (Sec.~\ref{subsec:recursive}). Finally, the results are discussed in Sec.~\ref{sec:conclusions}. Two appendices are included. Appendix~\ref{sec:Hamiltonian} collects the expressions of the constraints of the system before gauge fixing. Appendix~\ref{sec:gauge} describes an equivalent quantization in terms of other gauge-invariant variables.

\section{Classical system}\label{sec:classical}

In the derivation of the (symplectic) canonical structure and the Hamiltonian constraint of the system, we essentially adapt the treatment of Halliwell and Hawking~\cite{HH85} to the case of flat, compact spatial sections. Thus, we admit the existence of a global foliation of the spacetime, parametrized by a time function $t$ (which we use as the time coordinate), and write the metric in Arnowitt-Deser-Misner (ADM) form, i.e., in terms of the lapse function $N$, the shift vector $N^i$ (or the covector $N_i$), and the three-metric of the spatial slices, $h_{ij}$. Here, the spatial indices $i,j$ run from 1 to 3. In a homogeneous and isotropic spacetime, the above quantities can be described just by a homogeneous lapse $N_0$, the logarithm of the scale factor (of the spatial metric) $\alpha$, and a static reference three-metric $^0h_{ij}$. Here we choose $^0h_{ij}$ as the standard flat metric on the three-torus $T^3$, with periodicity equal to $l_0$ in each of the orthonormal directions, for which we use the angular coordinates $\theta_i\in [0, l_0)$. The shift vector vanishes and neither $\alpha$ nor $N_0$ depend on the position. A fully inhomogeneous metric can then be constructed by adding variables that depend on time and on the space point. It is extremely convenient to expand these variables in the eigenbases of (scalar, vector, and tensor) harmonics provided by the Laplace-Beltrami (LB) operator of the reference three-metric. In the case under consideration, we introduce the \emph{real} eigenfunctions
\begin{equation}
\tilde Q_{\vec n,+} = \sqrt 2\cos\left(\frac{2\pi}{l_0}\vec n\cdot\vec\theta\right),\quad  \tilde Q_{\vec n,-} = \sqrt 2\sin\left(\frac{2\pi}{l_0}\vec n\cdot\vec\theta\right),
\end{equation}
where $\vec n=(n_1,n_2,n_3)\in\mathbb Z^3$ and its first nonvanishing component is, e.g., strictly positive. We use the standard notation $\vec n\cdot\vec\theta=\sum_in_i\theta_i$. Notice that this basis of scalar modes is normalized so that
\begin{equation}
\int_{T^3}\!\!d^3\theta\,\tilde Q_{\vec n,\epsilon}(\vec\theta)\tilde Q_{\vec n',\epsilon'}(\vec\theta) = l_0^3\,\delta_{\vec n,\vec n'}\delta_{\epsilon,\epsilon'},
\end{equation}
$l_0^3$ being the fiducial volume of the three-torus, and $\epsilon,\epsilon'=+,-$. The corresponding eigenvalue equation is
\begin{equation}
^0h^{ij}(\tilde Q_{\vec n,\epsilon})_{|ij} = -\omega_n^2\tilde Q_{\vec n,\epsilon},
\end{equation}
where the vertical bar denotes the (covariant) derivative and $\omega_n^2=4\pi^2\vec n\cdot\vec n/l_0^{2}$.

We can construct vector and tensor modes from these scalar ones by covariant differentiation. In this work, we include no other vector and tensor eigenfunctions of the LB operator, since they are anyway dynamically decoupled from the scalar ones at the considered perturbative order, and we will only focus on the study of scalar perturbations~\cite{Bardeen80}. Using therefore only scalar harmonics, we write the ADM decomposition of the metric in the following way:
\begin{subequations}\label{eqs:expansions}
\begin{align}
h_{ij}(t,\vec\theta) &= \big(\sigma e^{\alpha(t)}\big)^2\;{}^0h_{ij}(\vec\theta)\Big(1+2\sum_{\vec n,\epsilon}a_{\vec n,\epsilon} (t)\tilde Q_{\vec n,\epsilon}(\vec\theta)\Big) \nonumber\\
&\quad +6\big(\sigma e^{\alpha(t)}\big)^2\sum_{\vec n,\epsilon}b_{\vec n,\epsilon}(t)\left(\frac1{\omega_n ^2}(\tilde Q_{\vec n,\epsilon})_{|ij}(\vec\theta)+\frac13{}^0h_{ij}(\vec\theta)\tilde Q_{\vec n,\epsilon}(\vec\theta)\right), \\
N(t,\vec\theta) &= \sigma N_0(t)\Big(1+\sum_{\vec n,\epsilon}g_{\vec n,\epsilon}(t)\tilde Q_{\vec n,\epsilon}(\vec\theta)\Big), \\
N_i(t,\vec\theta) &= \sigma^2e^{\alpha(t)}\sum_{\vec n,\epsilon}\frac1{\omega_n^2}k_{\vec n,\epsilon}(t)(\tilde Q_{\vec n,\epsilon})_{|i}(\vec\theta),
\end{align}
with $\sigma^2=4\pi G/(3l_0^3)$, and $G$ denotes the Newton constant. Besides, in all the sums over the tuples $\vec n$, here and in the following, the zero mode $\vec n=(0,0,0)$ is {\emph {excluded}}. This mode is already accounted for by considering the homogeneous variables, where we include its contribution. In this way, we see that the time-dependent coefficients $a_{\vec n,\epsilon}$, $b_{\vec n,\epsilon}$, $g_{\vec n,\epsilon}$, and $k_{\vec n,\epsilon}$ parametrize the inhomogeneities. The matter content of the universe, given by a scalar field $\Phi$ of mass $m=\tilde m/\sigma$, can also be expanded in the same basis:
\begin{equation}
\Phi(t,\vec\theta) = \frac1{l_0^{3/2}\sigma}\Big(\varphi(t)+\sum_{\vec n,\epsilon}f_{\vec n,\epsilon}(t)\tilde Q_{\vec n,\epsilon} (\vec\theta)\Big).
\end{equation}
\end{subequations}
The variable $\varphi$ determines the homogeneous part of the field, while the inhomogeneities are codified by the coefficients $f_{\vec n,\epsilon}$ [again, $\vec n\neq(0,0,0)$].

Substituting expressions~\eqref{eqs:expansions} in the Einstein-Hilbert action, one obtains the Lagrangian of the system in terms of the new variables, adapted to the expansion in harmonics. However, at this point and in what follows, we will treat the inhomogeneities as perturbations around the homogeneous background. Thus, we will truncate the action at second order in the perturbative coefficients. The expression that we get in this way for the flat case differs from the action of the model on the three-sphere, derived in Ref.~\cite{HH85} (see also Ref.~\cite{FMOV12} for more details), essentially only by the terms arising from the three-curvature of the spatial sections. As in that case, the standard procedure leads to a Hamiltonian of the form
\begin{equation}\label{eq:Hamiltonian}
H = N_0\Big(H_{|0}+\sum_{\vec n,\epsilon}H^{\vec n,\epsilon}_{|2}\Big)+\sum_{\vec n,\epsilon}N_0g_{\vec n,\epsilon}H^{\vec n,\epsilon}_{|1}+\sum_{\vec n,\epsilon}k_{\vec n,\epsilon}H^{\vec n,\epsilon}_{\_1},
\end{equation}
which is a linear combination of constraints. Here
\begin{equation}\label{eq:H_0}
H_{|0} = \tfrac12e^{-3\alpha}\big(-\pi_\alpha^2+\pi_\varphi^2+e^{6\alpha}\tilde m^2\varphi^2),
\end{equation}
while the explicit expressions of $H^{\vec n,\epsilon}_{|2}$, $H^{\vec n,\epsilon}_{|1}$, and $H^{\vec n,\epsilon}_{\_1}$ can be found in Appendix~\ref{sec:Hamiltonian}. We have called $\pi_q$ the momentum canonically conjugate to the generic variable $q$.

Therefore, the homogeneous Hamiltonian constraint $H_{|0}$ is corrected with second-order terms $H^{\vec n,\epsilon}_{|2}$ [with an independent contribution of each of the modes (${\vec n,\epsilon}$)]. The homogeneous lapse $N_0$ appears in the Hamiltonian as the Lagrange multiplier of the corrected Hamiltonian constraint. The other Lagrange multipliers, $N_0g_{\vec n,\epsilon}$ and $k_{\vec n,\epsilon}$, are related to first-order constraints: the linear Hamiltonian one $H^{\vec n,\epsilon}_{|1}$ and the diffeomorphism constraint $H^{\vec n,\epsilon}_{\_1}$, respectively. Their mode dependence indicates that these linear constraints are local in nature. In the two following subsections we propose two different gauge choices to fix the freedom associated with these local constraints.

\subsection{Gauge of flat spatial sections}

We first explore the gauge in which the spatial slices have constant curvature, i.e., the gauge in which
\begin{equation}\label{eqs:gaugeA}
a_{\vec n,\epsilon}= 0 = b_{\vec n,\epsilon},
\end{equation}
and hence $h_{ij} = (\sigma e^\alpha)^2\;{}^0h_{ij}$. Let us notice that this choice of gauge does not alter the symplectic structure of the remaining variables after reduction of the system. The above conditions are of second order with respect to the constraints and, what is more, they are well posed away from the section of the phase space where $\pi_\alpha=0$. The reduction of the system to the phase-space hypersurface defined by those conditions can be performed as in Ref.~\cite{FMOV12}, restraining the value of the canonical momenta of the fixed variables, $\pi_{a_{\vec n,\epsilon}}$ and $\pi_{b_{\vec n,\epsilon}}$, by solving the linear constraints after imposing Eq. \eqref{eqs:gaugeA}. In addition, the demand of the dynamical consistency of the gauge-fixing conditions (i.e., of the vanishing of their Poisson brackets with the total Hamiltonian), fixes the value of the Lagrange multipliers $k_{\vec n,\epsilon}$ and $N_0g_{\vec n,\epsilon}$. After the reduction, only one (homogeneous) constraint is left:
\begin{equation}\label{eq:reducedH}
H = N_0\Big(H_{|0}+\sum_{\vec n,\epsilon}H^{\vec n,\epsilon}_{|2}\Big),
\end{equation}
where the second-order Hamiltonian has the structure
\begin{equation}\label{eq:reducedH^n_|2}
H^{\vec n,\epsilon}_{|2} = \frac12e^{-\alpha}\big(E^n_{\pi\pi}\pi_{f_{\vec n,\epsilon}}^2+2E^n_{f\pi}f_{\vec n,\epsilon}\pi_{f_{\vec n,\epsilon}}+E^n_{ff}f_{\vec n,\epsilon}^2\big).
\end{equation}
In this quadratic expression in terms of the configuration and momentum variables of the mode expansion of the scalar field, the $E^n$-coefficients are given by
\begin{subequations}
\begin{align}
E^n_{\pi\pi} &= e^{-2\alpha}, \\
E^n_{f\pi} &= -3e^{-2\alpha}\frac{\pi_\varphi^2}{\pi_\alpha}, \\
E^n_{ff} &= \omega_n^2e^{2\alpha}+\tilde m^2e^{4\alpha}+3e^{-2\alpha}\left(3\pi_\varphi^2-2e^{6\alpha}\tilde m^2\varphi\frac{\pi_\varphi}{\pi_\alpha}\right).
\end{align}
\end{subequations}
Unlike in a closed universe~\cite{FMOV12}, these coefficients have no subdominant terms of the order of $\omega_n^{-2}$. This greatly simplifies the subsequent treatment.

The second-order Hamiltonian $H^{\vec n,\epsilon}_{|2}$, which by definition is quadratic in the perturbative variables, can be put in a KG form by means of a canonical transformation that respects the linearity of (the symplectic structure and the space of solutions for) the inhomogeneous sector. Remarkably, such a transformation involves the scaling of the field as a whole by a function of the background variables---namely the scale factor. Of course, the inverse scaling must be applied to the conjugate momentum in order to preserve the symplectic structure, but it is also necessary to add a term proportional to the field configuration to remove the cross-terms that couple field configuration and momentum in the Hamiltonian. The new variables are
\begin{subequations}\label{eqs:newvariablesA}
\begin{align}
\bar f_{\vec n,\epsilon} &= e^\alpha f_{\vec n,\epsilon}, \\
\pi_{\bar f_{\vec n,\epsilon}} &= e^{-\alpha}\left[\pi_{f_{\vec n,\epsilon}}-\left(3\frac{\pi_\varphi^2}{\pi_\alpha}+\pi_\alpha\right)f_{\vec n,\epsilon}\right], \\
\bar\alpha &= \alpha-\frac12\left(3\frac{\pi_\varphi^2}{\pi_\alpha^2}-1\right)\sum_{\vec n,\epsilon}f_{\vec n,\epsilon}^2, \\
\pi_{\bar\alpha} &= \pi_\alpha-\sum_{\vec n,\epsilon}\left[f_{\vec n,\epsilon}\pi_{f_{\vec n,\epsilon}}-\left(3\frac{\pi_\varphi^2}{\pi_\alpha}+\pi_\alpha\right)f_{\vec n,\epsilon}^2\right], \\
\bar\varphi &= \varphi+3\frac{\pi_\varphi}{\pi_\alpha}\sum_{\vec n,\epsilon}f_{\vec n,\epsilon}^2, \\
\pi_{\bar\varphi} &= \pi_\varphi.
\end{align}
\end{subequations}
This transformation is analogous to that performed in Ref.~\cite{FMOV12} for the case of a universe with positive-curvature spatial sections. Note that the homogeneous variables are corrected with second-order terms, which do not affect the zeroth-order Hamiltonian, but contribute to the second-order one (higher-order corrections are neglected). Thus, $\bar H_{|0}$ is obtained from $H_{|0}$~\eqref{eq:H_0} by just replacing the original variables with the barred ones. In turn, the second-order Hamiltonian adopts the form
\begin{equation}\label{eq:KGHamiltonian}
\bar H^{\vec n,\epsilon}_{|2} = \frac12e^{-\bar\alpha}\big(\pi_{\bar f_{\vec n,\epsilon}}^2+\bar E^n_{ff} {\bar f}_{\vec n,\epsilon}^2\big),
\end{equation}
with
\begin{align}
\bar E^n_{ff} &= \omega_n^2+\tilde m^2e^{2\bar\alpha}+\frac12e^{-4\bar\alpha}\big(-\pi_{\bar\alpha}^2+30\pi_{\bar\varphi}^2-3e^{6\bar\alpha}\tilde m^2\bar\varphi^2\big) \nonumber\\
&\quad -\frac92e^{-4\bar\alpha}\frac{\pi_{\bar\varphi}^2}{\pi_{\bar\alpha}^2}\big(3\pi_{\bar\varphi}^2-e^{6\bar\alpha}\tilde m^2\bar\varphi^2\big)-12e^{2\bar\alpha}\tilde m^2\bar\varphi\frac{\pi_{\bar\varphi}}{\pi_{\bar\alpha}}.
\end{align}
Therefore, the scaled field obeys a KG equation with time-dependent mass. From Eq.~\eqref{eq:KGHamiltonian}, and ignoring higher-order corrections,
\begin{equation}\label{eq:KGeq}
\ddot{\bar f}_{\vec n,\epsilon}+\bar E^n_{ff} \bar f_{\vec n,\epsilon} = 0,
\end{equation}
where the overdot stands for the derivative with respect to the conformal time $\eta=\int N_0e^{-\alpha}dt$. Note that it is irrelevant to use the barred or the unbarred homogeneous variables at the considered perturbative order. Taking into account the equations of motion of the background and the homogeneous Hamiltonian constraint, Eq.~\eqref{eq:KGeq} can be rewritten as
\begin{equation}\label{eq:MSequation}
\ddot{\bar f}_{\vec n,\epsilon} + \left(\omega_n^2-\frac{\ddot z}{z}\right)\bar f_{\vec n,\epsilon} = 0,
\end{equation}
with $z=-e^{\bar \alpha}\pi_{\bar\varphi}/\pi_{\bar \alpha}$. Of course, this is the well-known MS equation~\cite{Mukhanov05}. This should not be surprising since, in the chosen gauge, $\bar f_{\vec n,\epsilon}$ coincides with the modes of the MS variable $v_{\vec n,\epsilon}$, whose expression in a general gauge is
\begin{equation}
v_{\vec n,\epsilon} = e^\alpha\left[f_{\vec n,\epsilon}+\frac{\pi_{\varphi}}{\pi_\alpha}(a_{\vec n,\epsilon}+b_{\vec n,\epsilon})\right],
\end{equation}
in the notation used here
(see also the discussion about gauge-invariant quantities for the closed case in Appendix C of Ref.~\cite{FMO12}). In the flat case, the widely used, gauge-invariant MS variable satisfies the KG-type equation \eqref{eq:MSequation}. This, together with the relation $\dot{\bar f}_{\vec n,\epsilon}=\pi_{\bar f_{\vec n,\epsilon}}$ (arising from our choice of field momentum), allows us to apply straightforwardly the uniqueness results for the quantization of a scalar field with time-dependent quadratic potential in a three-torus~\cite{CCMMV12}. So, we know that there is a unique class of unitarily equivalent Fock representations for the scaled field with an invariant vacuum and unitarily implementable field dynamics in the corresponding background. A representative of this class is characterized by the complex structure\footnote{A complex structure is a real, linear map on the phase space that preserves the symplectic structure, and whose square is minus the identity~\cite{QFTCS}.} determined by the choice of annihilation-like variables
\begin{equation}\label{eq:annihilation}
a_{\bar f_{\vec n,\epsilon}} = \frac1{\sqrt{2\omega_n}}(\omega_n\bar f_{\vec n,\epsilon}+i\pi_{\bar f_{\vec n,\epsilon}})
\end{equation}
and the corresponding creation-like variables $(a_{\bar f_{\vec n,\epsilon}})^*$ (here, the star symbol denotes complex conjugation), which would be naturally associated with the massless situation. Moreover, the requirements of invariance and unitary dynamics cannot be satisfied if a different global scaling of the field is chosen.

Leaving aside the variables $\bar f_{\vec n,\epsilon}$, that determine the inhomogeneous modes of the matter field, the physical interpretation of the barred variables~\eqref{eqs:newvariablesA} in terms of the metric is not straightforward. Retaining only linear contributions of the perturbations---since we have disregarded second and higher-order perturbations in the nonzero modes of the metric variables in our analysis, because they do not affect the perturbative truncation of the action at quadratic order---, we obtain $h_{ij} = (\sigma e^{\bar\alpha})^2\;{}^0h_{ij}$ and
\begin{subequations}
\begin{align}
N &= \sigma N_0\left(1+3e^{-\bar\alpha}\frac{\pi_{\bar\varphi}}{\pi_{\bar\alpha}}\sum_{\vec n,\epsilon}\bar f_{\vec n,\epsilon}\tilde Q_{\vec n,\epsilon}\right), \\
N_i &= -\sigma^2\frac{N_0}{\pi_{\bar\alpha}}\sum_{\vec n,\epsilon}\frac3{\omega_n^2}\left[\pi_{\bar f_{\vec n,\epsilon}}+e^{-2\bar\alpha}\left(3\frac{\pi_{\bar\varphi}^2}{\pi_{\bar\alpha}}-2\pi_{\bar\alpha}\pi_{\bar\varphi}+e^{6\bar\alpha}\tilde m^2\bar\varphi\right)\bar f_{\vec n,\epsilon}\right](\tilde Q_{\vec n,\epsilon})_{,i}, \\
\Phi &= \frac1{l_0^{3/2}\sigma}\left({\bar \varphi}+e^{-\bar\alpha}\sum_{\vec n,\epsilon}\bar f_{\vec n,\epsilon}\tilde Q_{\vec n,\epsilon}\right).
\end{align}
\end{subequations}

\subsection{Longitudinal gauge}

The longitudinal gauge, in which the three-metric is conformally flat and the shift vector is zero, is employed frequently in the literature. Since $k_{\vec n,\epsilon}$ is just a Lagrange multiplier, $k_{\vec n,\epsilon}=0$ cannot be used as a gauge-fixing condition. The vanishing of the shift must be imposed with a suitably chosen restriction. As in Ref.~\cite{FMOV12}, the appropriate conditions are
\begin{equation}\label{eqs:gaugeB}
C_{\vec n,\epsilon} \equiv \pi_{a_{\vec n,\epsilon}}-\pi_\alpha a_{\vec n,\epsilon}-3\pi_\varphi f_{\vec n,\epsilon} = 0,\quad b_{\vec n,\epsilon} = 0.
\end{equation}
In the hypersurface defined by these equations, the constraint $H^{\vec n,\epsilon}_{\_1}=0$ amounts to demand that $\pi_{b_{\vec n,\epsilon}}=0$, whereas $H^{\vec n,\epsilon}_{|1}=0$ implies that
\begin{equation}\label{eq:a_n}
a_{\vec n,\epsilon} = 3\frac{\pi_\varphi\pi_{f_{\vec n,\epsilon}}+\big(e^{6\alpha} \tilde m^2\varphi-3\pi_\alpha\pi_\varphi\big)f_{\vec n,\epsilon}}{9\pi_\varphi^2+\omega_n^2e^{4\alpha}}.
\end{equation}
In order to obtain this expression, the constraints have been used and third-order terms have been neglected. On the other hand, the dynamical consistency of the conditions~\eqref{eqs:gaugeB} requires indeed the vanishing of the shift vector, as originally intended.

In this gauge, the nonzero value of the terms $\dot a_{\vec n,\epsilon}\pi_{a_{\vec n,\epsilon}}$ in the Lagrangian contribute to the action after the reduction of the system, even if the canonical pairs $(a_{\vec n,\epsilon},\pi_{a_{\vec n,\epsilon}})$ are eliminated as physical degrees of freedom. One can see that, as a consequence, the Poisson brackets of the remaining variables change. At this point, we introduce a new set of coordinates in which the reduced symplectic structure actually adopts a canonical form:
\begin{subequations}\label{eqs:newvariablesB}
\begin{align}
\bar f_{\vec n,\epsilon} &= e^\alpha f_{\vec n,\epsilon}, \\
\pi_{\bar f_{\vec n,\epsilon}} &= e^{-\alpha}(\pi_{f_{\vec n,\epsilon}}-3\pi_\varphi a_{\vec n,\epsilon}-\pi_\alpha f_{\vec n,\epsilon}), \\
\bar\alpha &= \alpha+\frac12\sum_{\vec n,\epsilon}\big(a_{\vec n,\epsilon}^2+f_{\vec n,\epsilon}^2\big), \\
\pi_{\bar\alpha} &= \pi_\alpha-\sum_{\vec n,\epsilon}\big(f_{\vec n,\epsilon}\pi_{f_{\vec n,\epsilon}}-3\pi_\varphi a_{\vec n,\epsilon}f_{\vec n,\epsilon}-\pi_\alpha f_{\vec n,\epsilon}^2\big), \\
\bar\varphi &= \varphi+3\sum_{\vec n,\epsilon}a_{\vec n,\epsilon}f_{\vec n,\epsilon}, \\
\pi_{\bar\varphi} &= \pi_\varphi.
\end{align}
\end{subequations}
These new variables are formally the same as those introduced in the case of perturbations around a closed FLRW model in the longitudinal gauge~\cite{FMOV12}. Recall that, after the reduction, $a_{\vec n,\epsilon}$ takes the value given in Eq.~\eqref{eq:a_n} (which does change slightly in the closed case). One can check that, in the reduction process, up to total time derivatives and neglecting fourth-order terms,
\begin{equation}
\dot\alpha\pi_\alpha+\dot\varphi\pi_{\varphi}+\sum_{\vec n,\epsilon}(\dot a_{\vec n,\epsilon}\pi_{a_{\vec n,\epsilon}}+\dot f_{\vec n,\epsilon}\pi_{f_{\vec n,\epsilon}}) = \dot{\bar\alpha}\pi_{\bar\alpha}+\dot{\bar\varphi}\pi_{\bar\varphi}+\sum_{\vec n,\epsilon}\dot{\bar f}_{\vec n,\epsilon}\pi_{\bar f_{\vec n,\epsilon}}.
\end{equation}
Hence, the barred variables have canonical (strictly speaking Dirac) brackets.

Note that we have taken advantage of the change of variables to also scale the matter field perturbation by the background scale factor. The choice of the conjugate momentum has been made following criteria similar to those of the previous subsection. In this case, we cannot remove completely the cross-terms of the second-order Hamiltonian with a \emph{local} canonical transformation, linear in the inhomogeneous sector.\footnote{The cross-terms could still be removed by transforming each mode differently. However, that would break the locality of the formalism. In any case, these kinds of transformations do not spoil the uniqueness results~\cite{FMO12,CFMM13}. } Nevertheless, this choice of momentum makes such cross-terms subdominant in the large-$\omega_n$ limit, as we will see below. In fact, the reduced Hamiltonian has again the structure given by Eqs.~\eqref{eq:reducedH} and \eqref{eq:reducedH^n_|2}, where the zeroth-order Hamiltonian can be obtained from Eq.~\eqref{eq:H_0} by just replacing the original variables with the new ones, while the coefficients of the second-order Hamiltonian are
\begin{subequations}
\begin{align}
\bar E^n_{\pi\pi} &= 1-\frac3{\omega_n^2}e^{-4\bar\alpha}\pi_{\bar\varphi}^2, \\
\bar E^n_{f\pi} &= -\frac3{\omega_n^2}e^{-6{\bar\alpha}}\pi_{\bar\varphi}\big(e^{6\bar\alpha}\tilde m^2\bar\varphi-2\pi_{\bar\alpha}\pi_{\bar\varphi}\big), \\
\bar E^n_{ff} &= \omega_n^2+\tilde m^2e^{2\bar\alpha}-\frac12e^{-4\bar\alpha}\big(\pi_{\bar\alpha}^2+15\pi_{\bar\varphi}^2+3e^{6\bar\alpha}\tilde m^2\bar\varphi^2\big) \nonumber\\
&\quad -\frac3{\omega_n^2}e^{-8\bar\alpha}\big(e^{6\bar\alpha}\tilde m^2\bar\varphi-2\pi_{\bar\alpha}\pi_{\bar\varphi}\big)^2.
\end{align}
\end{subequations}
We see that, indeed, $\bar E^n_{f\pi}$ is of the order of $\omega_n^{-2}$, and that $\bar E^n_{\pi\pi}$ and $\bar E^n_{ff}$ have subdominant terms of the same order (as in a closed universe~\cite{FMOV12}, were it not for the fact that the explicit expressions are slightly different). Consequently, the equation of motion for $\bar f_{\vec n,\epsilon}$ is not exactly of the KG type~\eqref{eq:KGeq}. Nonetheless, the results of uniqueness for the quantization of a KG field~\cite{CMOV11,CMOV12,CCMMV12} can be easily extended to include this very case, in the same way as Ref.~\cite{FMOV12} does for the case of scalar perturbations around a closed FLRW model.

More specifically, the equation of motion for the mode $\bar f_{\vec n,\epsilon}$ can be written in the form
\begin{equation}\label{eq:quasiKGeq}
\ddot{\bar f}_{\vec n,\epsilon}+r_n\dot{\bar f}_{\vec n,\epsilon}+(\omega_n^2+s+s_n)\bar f_{\vec n,\epsilon} = 0,
\end{equation}
where $s$ is the time-dependent mass
\begin{equation}
s = \tilde m^2e^{2\bar\alpha}-\frac12e^{-4\bar\alpha}\big(\pi_{\bar\alpha}^2+21\pi_{\bar\varphi}^2+3e^{6\bar\alpha}\tilde m^2\bar\varphi^2\big),
\end{equation}
while $r_n$ and $s_n$ are subdominant time-dependent functions that decay as $\omega_n^{-2}$ in the limit of infinitely large $\omega_n$. These quantities depend on time only through the homogeneous variables, and on the mode only through $\omega_n$.

On the other hand, the momentum canonically conjugate to ${\bar f}_{\vec n,\epsilon}$ is
\begin{equation}
\pi_{\bar f_{\vec n,\epsilon}} = (1+p_n)\dot{\bar f}_{\vec n,\epsilon}+q_n\bar f_{\vec n,\epsilon},
\end{equation}
where $p_n$ and $q_n$ are time-dependent corrections (via their dependence on the homogeneous variables) of the order of $\omega_n^{-2}$. Thus, $\pi_{\bar f_{\vec n,\epsilon}}$ is not exactly the time derivative of the corresponding configuration mode, but has additional contributions. These contributions are sufficiently subdominant, thanks to the choice of the barred variables~\eqref{eqs:newvariablesB}.

It is clear that, strictly speaking, the uniqueness results for the quantization of a KG field with time-dependent quadratic potential in the three-torus~\cite{CCMMV12} do not hold now, owing to the appearance of the terms $p_n$, $q_n$, $r_n$, and $s_n$. However, let us define annihilation-like variables as in Eq.~\eqref{eq:annihilation}. Given the linearity of the field dynamics and the decoupling of the modes, their classical evolution from an arbitrary reference time $\eta_0$ is given by a symplectic transformation of the form
\begin{equation}\label{eq:dynamics}
a_{\bar f_{\vec n,\epsilon}}(\eta) = \alpha_n(\eta,\eta_0)a_{\bar f_{\vec n,\epsilon}}(\eta_0)+\beta_n(\eta,\eta_0)a_{\bar f_{\vec n,\epsilon}}^*(\eta_0),
\end{equation}
where the Bogoliubov coefficients $\alpha_n$ and $\beta_n$ satisfy $|\alpha_n|^2-|\beta_n|^2=1$ at any time, and of course can be determined explicitly if the general solution to Eq.~\eqref{eq:quasiKGeq} is known. Nevertheless, for our purposes, we do not need the exact expressions of these quantities---it suffices to know their asymptotic behavior in the large-$\omega_n$ limit, which is not essentially altered by the corrections $p_n$, $q_n$, $r_n$, and $s_n$~\cite{FMOV12, CCMMV12}:
\begin{equation}
\alpha_n(\eta,\eta_0) = e^{-i\omega_n(\eta-\eta_0)}+O(\omega_n),\quad \beta_n(\eta,\eta_0) = O\big(\omega_n^{-2}\big).
\end{equation}
The symbol $O$ denotes the asymptotic order. Since the proof of the aforementioned uniqueness rests mainly on the asymptotic behavior of the coefficients $\alpha_n$ and $\beta_n$, it can be easily extended to our case with subdominant corrections. Thus, the annihilation-like variables defined just as in Eq.~\eqref{eq:annihilation} determine a complex structure that, apart from being invariant under translations in the three-torus, permits a quantum unitary implementation of the classical evolution~\eqref{eq:dynamics}. The corresponding Fock representation belongs to a unitary equivalence class that contains all the representations with those properties. Furthermore, such an equivalence class ceases to exist for any other, different scaling of the field or choice of momentum, so the election of the barred variables~\eqref{eqs:newvariablesB} is key for this result on the translation invariance and unitary dynamics of the Fock quantization.

The necessary and sufficient condition for the unitary implementation of the field dynamics is that the sum $\sum_{\vec n,\epsilon}|\beta_n|^2$ must converge~\cite{hs}. If we call $G_n$ the degeneracy of each eigenvalue $-\omega_n^2$ of the LB operator, the previous sum can be rearranged as $\sum_n G_n|\beta_n|^2$. Now, although the dependence of $G_n$ on the eigenvalue is very complicated (because of the possible accidental degeneracy beyond that in permutations in the tuple $\vec n$ and flip of signs in its components), the sum does indeed converge, since $G_n$ can be bounded from above by a quantity that grows asymptotically as $\omega_n^2$~\cite{FMOV12}. Hence, there exists a unitary operator that implements the dynamical transformations~\eqref{eq:dynamics} at the quantum level, i.e., in the Fock space of the chosen representation. The uniqueness of this representation is proven along similar lines.

As we have seen in the previous subsection, the MS variable defines a privileged representation with unitary dynamics and invariant complex structure, as it is canonically conjugate to its time derivative and it satisfies a KG equation with time-dependent mass. These properties hold independently of the gauge \emph{in the case of a flat universe}. However, in the longitudinal gauge, the scaled perturbation of the matter field does not coincide with the MS variable and, since the relation between the two variables is mode dependent, the question may be raised as to whether the corresponding preferred quantizations are unitarily equivalent. In spite of this concern, it turns out that the answer to this question is in fact in the affirmative, in the light of a recent work~\cite{CFMM13} that addresses the unitary implementability of these kinds of time and mode-dependent linear canonical transformations which do not mix different modes of the LB operator nor alter significantly the field equations (namely, they only affect the KG form at most by introducing innocuous subdominant terms in the ultraviolet limit). These transformations have been proven to be unitary in the quantum theory~\cite{CFMM13}, thus reinforcing even more the results of quantization uniqueness.

Specifically, the transformation to the MS variable and its conjugate momentum has the form
\begin{subequations}\label{eqs:MStrans}
\begin{align}
v_{\vec n,\epsilon} &= A_n\bar f_{\vec n,\epsilon}+B_n\pi_{\bar f_{\vec n,\epsilon}}, \\
\pi_{v_{\vec n,\epsilon}} = \dot v_{\vec n,\epsilon} &= C_n\bar f_{\vec n,\epsilon}+D_n\pi_{\bar f_{\vec n,\epsilon}}.
\end{align}
The coefficients of this linear transformation are functions of the homogeneous variables:
\end{subequations}
\begin{subequations}
\begin{align}
A_n &= 1+\frac3{\omega_n^2}e^{-4\bar\alpha}\frac{\pi_{\bar\varphi}}{\pi_{\bar\alpha}}\big(e^{6\bar\alpha}\tilde m^2\bar\varphi-2\pi_{\bar\alpha}\pi_{\bar\varphi}\big), \\
B_n &= \frac3{\omega_n^2}e^{-2\bar\alpha}\frac{\pi_{\bar\varphi}^2}{\pi_{\bar\alpha}}, \\
C_n &= -3e^{-2\bar\alpha}\frac{\pi_{\bar\varphi}^2}{\pi_{\bar\alpha}}-\frac3{\omega_n^2}e^{-6\bar\alpha}\frac1{\pi_{\bar\alpha}}\bigg[e^{12\bar\alpha}\tilde m^4\bar\varphi^2+\frac12\pi_{\bar\varphi}^2\big(11\pi_{\bar\alpha}^2-15\pi_{\bar\varphi}^2-3e^{6\bar\alpha}\tilde m^2\bar\varphi^2\big)\bigg] \nonumber\\
&\quad +\frac{3}{2\omega_n^2} \tilde m^2\bar\varphi\frac{\pi_{\bar\varphi}}{\pi_{\bar\alpha}^2}\big(5\pi_{\bar\alpha}^2-3\pi_{\bar\varphi}^2+3e^{6\bar\alpha}\tilde m^2\bar\varphi^2\big), \\
D_n &= 1-\frac3{2\omega_n^2}e^{-4\bar\alpha}\frac{\pi_{\bar\varphi}}{\pi_{\bar\alpha}}\bigg[2e^{6\bar\alpha}\tilde m^2\bar\varphi-\frac{\pi_{\bar\varphi}}{\pi_{\bar\alpha}}\big(\pi_{\bar\alpha}^2-3\pi_{\bar\varphi}^2+3e^{6\bar\alpha}\tilde m^2{\bar \varphi}^2)\bigg].
\end{align}
\end{subequations}
As anticipated, this transformation depends on the mode through $\omega_n$, making it nonlocal. By employing the Hamiltonian constraint, one can check that $A_nD_n-B_nC_n=1$ up to the linear perturbative order. Consequently, the transformation is in fact canonical, as far as the inhomogeneous variables are concerned. It can be completed to a genuine canonical transformation on the whole of the phase space of the reduced system by introducing new homogeneous variables, differing from the previous ones by additional quadratic terms in the perturbations. We will not provide the explicit expressions of those variables here, since they are complicated and we will not need them in the rest of our discussion.

The privileged quantization of the inhomogeneities corresponding to the gauge-invariant quantities $v_{\vec n,\epsilon}$ and $\pi_{v_{\vec n,\epsilon}}$ can be constructed from the following choice of annihilation-like variables:
\begin{equation}
a_{v_{\vec n,\epsilon}} = \frac1{\sqrt{2\omega_n}}(\omega_nv_{\vec n,\epsilon}+i\pi_{v_{\vec n,\epsilon}}).
\end{equation}
Now, written in terms of annihilation and creation-like variables, transformation~\eqref{eqs:MStrans} is of Bogoliubov type,
\begin{equation}
a_{v_{\vec n,\epsilon}} = \lambda_n^+a_{\bar f_{\vec n,\epsilon}}+\lambda_n^-a_{\bar f_{\vec n,\epsilon}}^*,
\end{equation}
with Bogoliubov coefficients
\begin{equation}
\lambda_n^\pm = \frac12(A_n\pm D_n)+\frac i{2\omega_n}(C_n\mp\omega_n^2B_n),
\end{equation}
such that $|\lambda_n^+|^2-|\lambda_n^-|^2=1$. The ``beta coefficients'' (i.e., the antilinear Bogoliubov coefficients)
$\lambda_n^-$ decrease as $\omega_n^{-2}$ in the large-$\omega_n$ limit. Therefore, the sum $\sum_n G_n|\lambda_n^-|^2$ converges, and the above transformation is unitarily implementable in the Fock quantization adopted for the inhomogeneities (regarding the homogeneous variables as background ones). Accordingly, even if in the longitudinal gauge we have two preferred Fock quantizations for the inhomogeneities, they are completely equivalent. In this sense, whether one keeps the scaled field or uses the MS variable is of no physical relevance. Note that, together with the results of the previous subsection, this also implies the unitary equivalence of the Fock quantization attained  for the inhomogeneous sector in the longitudinal gauge and in the gauge of flat spatial sections, therefore providing robustness to the physical consequences of the quantization beyond the specific gauge fixing adopted.

Finally, for completeness, let us write the metric in terms of the barred variables \eqref{eqs:newvariablesB}. Including only linear contributions of the perturbations, we get
\begin{subequations}
\begin{align}
h_{ij} &= \big(\sigma e^{\bar\alpha}\big)^2\;{}^0h_{ij}\Big(1+2\sum_{\vec n,\epsilon} a_{\vec n,\epsilon}\tilde Q_{\vec n,\epsilon}\Big), \\
N &= \sigma N_0\Big(1-\sum_{\vec n,\epsilon}a_{\vec n,\epsilon}\tilde Q_{\vec n,\epsilon}\Big), \\
N_i &= 0, \\
\Phi &= \frac1{l_0^{3/2}\sigma}\Big({\bar \varphi}+e^{-\bar\alpha}\sum_{\vec n,\epsilon}\bar f_{\vec n,\epsilon}\tilde Q_{\vec n,\epsilon}\Big),
\end{align}
\end{subequations}
where
\begin{equation}
a_{\vec n,\epsilon} = \frac3{\omega_n^2}e^{-3\bar\alpha}\big[\pi_{\bar\varphi} \pi_{\bar f_{\vec n,\epsilon}}+e^{-2\bar\alpha}\big(e^{6\bar\alpha}\tilde m^2\bar\varphi-2\pi_{\bar\alpha}\pi_{\bar\varphi}\big){\bar f}_{\vec n,\epsilon}\big].
\end{equation}

\section{Kinematical Hilbert space}\label{sec:quantization}

In this section, we proceed to the complete quantization of the model. With this aim, we introduce a kinematical Hilbert space where the Hamiltonian constraint can be represented. As anticipated, we adopt the polymeric quantization for the homogeneous degrees of freedom of the gravitational field, whereas we employ a standard Fock representation for the inhomogeneities.

\subsection{Homogeneous sector}\label{subsec:hom}

It is well known that a standard quantization of the homogeneous sector would not generally avoid the big-bang singularity that arises in classical General Relativity. This problem can be overcome with a polymeric quantization. However, the parametrization of the homogeneous sector introduced in Sec.~\ref{sec:classical} is not adapted to this kind of quantization, which starts from the spatial smearing of an Ashtekar-Barbero connection $A^a_i$ (in terms of holonomies) and of a densitized triad $E_a^i$ (via fluxes trough surfaces)~\cite{LQG}. Here, while $i$ is a spatial index as before, $a$ is an internal $\mathfrak{su}(2)$-index. In the homogeneous and isotropic case, the diffeomorphism and the Gauss constraints can be fixed so that these variables take the form~\cite{LQC}
\begin{equation}
A^a_i = c\frac{\,^0\omega^a_i}{l_0},\quad E_a^i = p\sqrt{^0h}\frac{\,{}^0e_a^i}{l_0^2},
\end{equation}
where $^0e_a^i$ is a fiducial triad in $T^3$ and $^0\omega^a_i$ is the corresponding co-triad, so that $^0h_{ij}=\delta_{ab}{}^0\omega^a_i\,{}^0\omega^b_j$. The time-dependent variables $c$ and $p$ parametrize the homogeneous gravitational sector and satisfy $\{c,p\}=8\pi G\gamma/3$, $\gamma$ being the Immirzi parameter. The relation between these variables and the ones used in Sec.~\ref{sec:classical} is
\begin{subequations}\label{eq:homchange}
\begin{equation}
|p| = l_0^2\sigma^2e^{2\alpha},\quad pc = -\gamma l_0^3\sigma^2\pi_\alpha.
\end{equation}
The ambiguity in the sign of $p$ is related to the orientation of the triad and is of no practical significance here. As for the homogeneous part of the scalar field, it is convenient to rescale it in order to facilitate the comparison with the LQC literature:
\begin{equation}
\phi = \frac{\varphi}{l_0^{3/2}\sigma},\quad \pi_\phi = l_0^{3/2}\sigma\pi_\varphi.
\end{equation}
\end{subequations}
With the new parametrization of the homogeneous sector, the zeroth-order Hamiltonian constraint reads
\begin{equation}\label{eq:classC_0}
C_0\equiv \frac{16\pi G}\sigma H_{|0} = -\frac6{\gamma^2}\sqrt{|p|}\,c^2+\frac{8\pi G}{|p|^{3/2}}\big(\pi_\phi^2+ m^2|p|^3\phi^2\big).
\end{equation}
Naturally, the additional factor $16\pi G/\sigma$ can be absorbed with a redefinition of the homogeneous lapse $\sigma N_0$.

The fundamental variables in the polymeric quantization are the holonomies of the connection and the fluxes of the densitized triad. In the homogeneous and isotropic case, it suffices to consider (i) holonomies of $A^a_i$ along straight edges of length $\mu l_0$ in the fiducial directions, whose matrix elements can be reconstructed from the functions $N_\mu=\exp(i\mu c/2)$, and (ii) fluxes of $E_a^i$ through square surfaces orthogonal to the fiducial directions, which are just proportional to $p$~\cite{APS06b}. The quantities $N_\mu$ and $p$ can be represented as operators acting on the kinematical Hilbert space $\mathcal H_\mathrm{kin}^\mathrm{grav}=L^2(\mathbb R_\mathrm{B}, d\mu_\mathrm{B})$, i.e., the space of square-integrable functions in the Bohr compactification of the real line $\mathbb R$, with the corresponding Haar measure $d\mu_\mathrm{B}$~\cite{Velhinho07}. In the so-called \emph{improved dynamics} scheme~\cite{APS06c}, one considers edges of fiducial length $\bar\mu l_0$, related to the minimum nonzero
eigenvalue allowed for the area operator in LQG, $\Delta$, by the formula $\bar\mu=\sqrt{\Delta/p}$. The action of the operators $\hat N_{\bar\mu}$ is especially simple in the \emph{orthonormal} basis $\{|v\rangle\}_{v\in\mathbb R}$ such that
\begin{equation}
\hat p|v\rangle = \sgn(v)(2\pi\gamma G\hbar\sqrt\Delta|v|)^{2/3}|v\rangle,
\end{equation}
namely
\begin{equation}
\hat N_{\bar\mu}|v\rangle = |v+1\rangle.
\end{equation}
Here, $\hbar$ is the reduced Plack constant. Apart from an orientation sign, the label $v$ is proportional to the eigenvalue of the volume operator $\hat V=|\hat p|^{3/2}$.

The polymeric quantization of the gravitational degrees of freedom is argued to capture the most significant effects of the discrete geometry, while for the homogeneous part of the scalar field we simply adopt a standard Schr\"odinger quantization, representing $\phi$ by the multiplication operator on $\mathcal H_\mathrm{kin}^\mathrm{matt}=L^2(\mathbb R,d\phi)$. The total homogeneous kinematical Hilbert space is therefore the product $\mathcal H_\mathrm{kin}^\mathrm{grav}\otimes\mathcal H_\mathrm{kin}^\mathrm{matt}$.

The representation of the Hamiltonian constraint of the homogeneous sector is constructed mimicking the strategy put forward in LQG for the full theory. The constraint is first written in terms of the elementary variables---the volume and the holonomies of the improved dynamics approach. In particular, the field strength of the connection is expressed in principle as the limit of a holonomy around a square loop as the enclosed area tends to zero. Then, this limit is replaced by fixing the area of the square to $\Delta$~\cite{APS06b,APS06c}, the ``area gap'' in LQG. Finally, the elementary variables are promoted to operators. We adopt here the so-called simplified MMO prescription~\cite{MOP11,MMO09}, in which a densitized version of the quantum constraint can be introduced via
\begin{equation}\label{eq:C_0}
\hat C_0 = \widehat{\left[\frac1{V}\right]}^{1/2}\hat{\mathcal C}_0\widehat{\left[\frac1{V}\right]}^{1/2}.
\end{equation}
The inverse-volume operator $(\widehat{1/V})=(\widehat{1/\sqrt{|p|}})^3$ is defined as the cube of the regularized operator
\begin{equation}
\widehat{\left[\frac1{\sqrt{|p|}}\right]} = \frac3{4\pi\gamma G\hbar\sqrt\Delta}\widehat{\sgn(p)}\sqrt{|\hat p|}\big(\hat N_{-\bar\mu}\sqrt{|\hat p|}\hat N_{\bar\mu}-\hat N_{\bar\mu}\sqrt{|\hat p|}\hat N_{-\bar\mu}\big),
\end{equation}
which is diagonal in the $v$-basis. This operator has a purely point spectrum. Remarkably, it is bounded, and its kernel is just the zero-volume state $|v=0\rangle$. On the other hand, $\hat{\mathcal C}_0$ adopts the form
\begin{equation}\label{eq:calC_0}
\hat{\mathcal C}_0 = -\frac6{\gamma^2}\hat\Omega_0^2+8\pi G\big({\hat \pi}_\phi^2+m^2\hat V^2\hat\phi^2\big),
\end{equation}
where the operator $\hat\Omega_0$ is given by
\begin{equation}\label{eq:Omega}
\hat\Omega_0 = \frac1{4i\sqrt\Delta}\hat V^{1/2}\big[\widehat{\sgn(v)}\big(\hat N_{2\bar\mu}-\hat N_{-2\bar\mu}\big)+\big(\hat N_{2\bar\mu}-\hat N_{-2\bar\mu}\big)\widehat{\sgn(v)}\big]\hat V^{1/2}.
\end{equation}
Even though the quantum constraint was not directly obtained from the classical one in the form given by Eq.~\eqref{eq:classC_0}, \emph{a posteriori} it seems easy to pass from one to the other with some simple substitutions. This will inspire the prescription that we will follow in Sec.~\ref{subsec:inhom} to quantize the part of the Hamiltonian constraint that is quadratic in the inhomogeneous modes.

The action of $\hat\Omega_0^2$ on an element of the $v$-basis is
\begin{equation}
\hat\Omega_0^2|v\rangle = -f_+(v)f_+(v+2)|v+4\rangle+\big[f_+^2(v)+f_-^2(v)\big]|v\rangle-f_-(v)f_-(v-2)|v-4\rangle,
\end{equation}
where
\begin{equation}
f_\pm(v) = \frac{\pi\gamma G\hbar}2\sqrt{|v|}\sqrt{|v\pm2|}[\sgn(v)+\sgn(v\pm2)].
\end{equation}
This second-order difference operator has some remarkable properties. Firstly, it leaves invariant the orthogonal complement of the zero-volume state, $\tilde{\mathcal H}_\mathrm{kin}^\mathrm{grav}$. This and the fact that $|v=0\rangle$ is annihilated by the inverse-volume operator in the constraint \eqref{eq:C_0}, allows us to restrict the subsequent analysis to the commented orthogonal complement, ignoring in practice $|v=0\rangle$, which decouples completely. In this sense, the singularity is \emph{resolved quantum mechanically}. Once the zero-volume state is removed, a bijection can be established between the (generalized) states annihilated by $\hat C_0$ and its densitized version $\hat{\mathcal C}_0$, of simpler form. Furthermore, the operator $\hat\Omega_0^2$ does not mix states with support on $v<0$ and $v>0$, owing to the combination of sign functions in $f_\pm$. Actually, it respects the Hilbert spaces $\mathcal H_\mathrm{\varepsilon}^\pm$ formed by states with support on the semilattices $\mathcal L_\mathrm{\varepsilon}^\pm=\{v=\pm(\varepsilon+4n)|n\in\mathbb N\}$, where $\varepsilon\in(0,4]$. The spaces $\mathcal H_\mathrm{\varepsilon}^\pm$ become therefore superselection sectors of the homogeneous model. Finally, the operator $\hat\Omega_0^2$ is self-adjoint in a suitable dense domain~\cite{MOP11, MMO09, KL08}. All these properties are inherited by the homogeneous Hamiltonian constraint $\hat{\mathcal C}_0$.

In the framework of our perturbative theory, we consider that all the operators defined in this section represent the homogeneous variables of our reduced model. Thus, up to a constant factor, the operator $\hat{C}_0$ will implement the zeroth-order part of the Hamiltonian constraint \emph{in the barred variables}. In other words, the classical variables $c$ and $p$ are defined as in Eqs.~\eqref{eq:homchange}, replacing the homogeneous variables with their barred counterparts~\eqref{eqs:newvariablesA} or \eqref{eqs:newvariablesB} (depending on the chosen gauge). Notice also that, in the part of the constraint that is quadratic in the inhomogeneities, this distinction between barred and unbarred variables is irrelevant up to the considered perturbative order.

\subsection{Inhomogeneous sector}\label{subsec:inhom}

For the inhomogeneous sector, we adopt a Fock quantization, following the hybrid approach. Quantum field theory in curved classical spacetimes is generally considered a physically meaningful approximation in a certain suitable regime. The ambiguity in the choice of a particular representation is circumvented by appealing to the uniqueness results considered in Sec.~\ref{sec:classical}. In this way, we expect to recover a unitary Fock quantization in the regime in which the behavior of the homogeneous degrees of freedom can be described by an effective background.

As explained in Sec.~\ref{sec:classical}, the requirements of unitary quantum field dynamics (in any finite interval of time, no matter how short) and of an invariant vacuum state under the translations of the three-torus select a preferred scaling of the perturbation, a unique conjugate field momentum, and a unique unitary equivalence class of Fock representations for their canonical commutation relations. A representative of this class is the massless representation, defined by the annihilation-like variables~\eqref{eq:annihilation} and their complex conjugate creation-like variables. Let us recall that, although the variables $\{\bar f_{\vec n,\epsilon}\}$ are just the modes of the matter-field perturbation scaled by the FLRW scale factor in the two gauge choices, the explicit expression of $\pi_{\bar f_{\vec n,\epsilon}}$ (and the barred homogeneous variables) in terms of the original variables depends on the gauge.

Given a complex structure (which is compatible with the symplectic structure and defines an inner product in the phase space~\cite{QFTCS}), the related Fock representation can be constructed by standard procedures. The associated annihilation-like variables (and their complex conjugate) are promoted to annihilation (and creation) operators $\hat a_{f_{\vec n,\epsilon}}$ ($\hat a_{f_{\vec n,\epsilon}}^\dagger$) acting in the usual way on the Fock space $\mathcal F$, formed by completion of the linear span of the orthonormal occupancy-number basis
\begin{equation}
\Big\{|\mathcal N\rangle=|N_{(1,0,0),+};N_{(1,0,0),-};\ldots\rangle\ \Big|\ N_{\vec n,\epsilon}\in\mathbb N,\sum_{\vec n,\epsilon}N_{\vec n,\epsilon}<\infty\Big\}.
\end{equation}
We take the total kinematical Hilbert space simply as the product of those of the homogeneous and the inhomogeneous sectors, $\mathcal H_\mathrm{kin}^\mathrm{tot} = \mathcal H_\mathrm{kin}^\mathrm{grav}\otimes\mathcal H_\mathrm{kin}^\mathrm{matt}\otimes\mathcal F$. The fundamental operators whose action has been defined only on one of the factors of the above product are promoted to act as the identity on the other pieces of the total kinematical Hilbert space. The action of the Hamiltonian constraint, on the contrary, will not respect the product structure of $\mathcal H_\mathrm{kin}^\mathrm{tot}$, since its part that is quadratic in the perturbations mixes the homogeneous and the inhomogeneous sectors, as we are about to see.

In principle, we do not have at our disposal a general procedure to regularize this quadratic part of the constraint, in contrast to the situation described for the homogeneous constraint. This is a crucial point because the classical variable $c$ has no quantum counterpart, for the polymeric representation fails to be continuous. To avoid this problem, we will follow a quantization prescription that draws inspiration from the accumulated experience in LQC. The basic idea is to promote the product $(cp)^2$ to the operator $\hat\Omega_0^2$. Then, any even power $(cp)^{2k}$ can be represented as $(\hat\Omega_0^2)^k$. As for the odd powers, the strategy must be changed, because $\hat\Omega_0$ [as defined in Eq.~\eqref{eq:Omega}] is a step-two difference operator, and hence it mixes different spaces $\mathcal H_\varepsilon^\pm$. Since we want the perturbed theory to respect the superselection sectors of the original, unperturbed theory, we introduce the step-4 difference operator
\begin{equation}
\hat\Lambda_0 = \frac1{8i\sqrt\Delta}\hat V^{1/2}\big[\widehat{\sgn(v)}\big(\hat N_{4\bar\mu}-\hat N_{-4\bar\mu}\big)+\big(\hat N_{4\bar\mu}-\hat N_{-4\bar\mu}\big)\widehat{\sgn(v)}\big]\hat V^{1/2},
\end{equation}
and use it to represent the even powers $(cp)^{2k+1}$ as $|\hat\Omega_0|^k\hat\Lambda_0|\hat\Omega_0|^k$. Actually, this strategy is similar to that adopted in the LQC description of FLRW universes to represent the Hubble parameter~\cite{MOP11}. Of course, there is still the usual ambiguity regarding the factor ordering. In this sense, for the product $\phi\pi_\phi$ we choose the symmetric operator $(\hat\phi\hat\pi_\phi+\hat\pi_\phi\hat\phi)/2$. Expressions involving the volume are symmetrized by splitting the corresponding power in two equal factors; e.g., a term like $V^kf(cp)$ (where $f$ is an arbitrary function) is promoted to the operator $\hat V^{k/2}\hat f\hat V^{k/2}$. Finally, in consonance with our choice of Fock representation, wherever products of annihilation and creation-like variables appear, we adopt normal ordering.

If we apply the above prescription to the purely homogeneous part of the constraint, we recover the operator $\hat C_0$ defined by Eqs.~\eqref{eq:C_0} and \eqref{eq:calC_0}. Of course, the total quantum constraint has the structure $\hat C=\hat C_0+\sum_{\vec n,\epsilon}\hat C^{\vec n,\epsilon}_2$, where the form of the terms that are quadratic in the inhomogeneous modes depends on the chosen gauge. Besides, for the complete constraint, it is possible to use exactly the same change of densitization that was introduced for the homogeneous sector in Eq.~\eqref{eq:C_0}. In the following subsections we provide the expressions that one obtains in this manner.

\subsubsection{Gauge of flat spatial sections}
In the gauge fixed by the conditions $a_{\vec n,\epsilon}=0=b_{\vec n,\epsilon}$, the quadratic part of the densitized constraint can be written as
\begin{equation}
\hat{\mathcal C}^{\vec n,\epsilon}_2 = \frac{8\pi G}{\tilde\omega_n}\widehat{\left[\frac1V\right]}^{-1/3}\left[\left(2{\tilde \omega}_n^2+\hat F\right)\hat N_{\vec n,\epsilon}+\frac12\hat F \hat{X}_{\vec n,\epsilon}^+\right]\widehat{\left[\frac1V\right]}^{-1/3},
\end{equation}
where ${\tilde \omega}_n=l_0\omega_n$, $\hat N_{\vec n,\epsilon}=\hat a_{\bar f_{\vec n,\epsilon}}^\dagger\hat a_{\bar f_{\vec n,\epsilon}}$, $\hat{X}_{\vec n,\epsilon}^+=\big(\hat a_{\bar f_{\vec n,\epsilon}}^\dagger\big)^2+(\hat a_{\bar f_{\vec n,\epsilon}}\big)^2$, and
\begin{align}
\hat F &=  m^2\hat V^{2/3}-\frac12\widehat{\left[\frac1V\right]}^{2/3}\left(\frac1{\gamma^2}\hat\Omega_0^2-40\pi G\hat\pi_\phi^2+48\gamma^2\pi^2G^2\hat\Omega_0^{-2}\hat\pi_\phi^4\right)\widehat{\left[\frac1V\right]}^{2/3} \nonumber\\
&\quad -2\pi Gm^2\hat V^{1/3}\left[\hat\phi^2-4\gamma|\hat\Omega_0|^{-1}\hat\Lambda_0|\hat\Omega_0|^{-1}(\hat\phi\hat\pi_\phi+\hat\pi_\phi\hat\phi)\right]\hat V^{1/3} \nonumber \\
&\quad + 2\gamma^2\pi^2G^2m^2\hat V^{1/3}\hat\Omega_0^{-2}\hat V^{1/3}(\hat\phi\hat\pi_\phi+\hat\pi_\phi\hat\phi)^2.
\end{align}

\subsubsection{Longitudinal gauge}

In the longitudinal gauge,
\begin{align}
\hat{\mathcal C}^{\vec n,\epsilon}_2 &= \frac{8\pi G}{\tilde\omega_n}\widehat{\left[\frac1V\right]}^{-1/3}\left(2\tilde\omega_n^2+\hat F_n^-\right)\widehat{\left[\frac1V\right]}^{-1/3}\hat N_{\vec n,\epsilon} \nonumber\\
& \quad +\frac{4\pi G}{\tilde\omega_n}\widehat{\left[\frac1V\right]}^{-1/3}\left(\hat F_n^+\hat{X}_{\vec n,\epsilon}^++i\frac{4\pi G}{{\tilde\omega}_n}\hat G \hat{X}_{\vec n,\epsilon}^-\right)\widehat{\left[\frac1V\right]}^{-1/3},
\end{align}
where ${\tilde\omega}_n$, $\hat N_{\vec n,\epsilon}$, and $\hat{X}_{\vec n,\epsilon}^+$ are defined as before, while $\hat{X}_{\vec n,\epsilon}^-=\big(\hat a_{\bar f_{\vec n,\epsilon}}^\dagger\big)^2-(\hat a_{\bar f_{\vec n,\epsilon}}\big)^2$, and
\begin{subequations}
\begin{align}
\hat F_n^\pm &=  m^2\hat V^{2/3}-\frac12\widehat{\left[\frac1V\right]}^{2/3}\left(\frac1{\gamma^2}\hat\Omega_0^2+4\pi G\big[(5\mp2)\hat\pi_\phi^2+m^2\hat V^2\hat\phi^2\big]\right)\widehat{\left[\frac1V\right]}^{2/3} \nonumber\\
&\quad -\frac{4\pi G}{\tilde\omega_n^2}\widehat{\left[\frac1V\right]}^{4/3}\left( \frac2\gamma\hat\Lambda_0\hat \pi_\phi+m^2\hat V^2\hat\phi\right)^2\widehat{\left[\frac1V\right]}^{4/3}, \\
\hat G &= -\widehat{\left[\frac1V\right]}\left(\frac4\gamma\hat\Lambda_0\hat \pi_\phi^2+m^2\hat V^2(\hat\phi \hat \pi_\phi+\hat \pi_\phi\hat\phi)\right)\widehat{\left[\frac1V\right]}.
\end{align}
Note that, as in the other gauge, the final expression can be cast in a form independent of $l_0$.
\end{subequations}

\section{Physical states}\label{sec:phys}

In this section, we present two different characterizations of the solutions to the quantum Hamiltonian constraint. The first one is based on the use of the homogeneous scalar field $\phi$ as a time. While this strategy is followed in LQC frequently, its simplicity is compromised when the scalar field has a nonzero mass. Besides, by recurring to it, one may get the impression that the validity of the quantum treatment rests heavily on the availability of a relational time of this kind. For this reason, an alternate characterization is presented in Sec.~\ref{subsec:recursive}, in which the solutions are characterized in terms of constant FLRW-volume sections, which the constraint relates recursively.

\subsection{Characterization in terms of a relational time}\label{subsec:time}

In a homogeneous and isotropic universe minimally coupled to an otherwise free, massless scalar field, the value of the field grows monotonically in every classical trajectory. Thus, the field can be interpreted as a global emergent time. However, such a simple model does not undergo (sufficient) inflation. The situation changes with the inclusion of a mass for the field: whereas inflation becomes possible, the field is no longer monotonic. Nevertheless, it can still be used as a relational time locally. On this basis, the constraint equation $(\Psi|\hat{\mathcal C}^\dagger=0$ can be interpreted as an evolution equation. Thus, expanding the physical states as
\begin{equation}
(\Psi|=\int_{-\infty}^\infty\!\!\!d\phi\sum_{v\in\mathcal L^{\pm}_{\varepsilon}}\sum_{\mathcal N}\langle\phi|\langle v|\langle\mathcal N|\Psi(\phi,v,\mathcal N),
\end{equation}
the wavefunction $\Psi$ must satisfy
\begin{equation}\label{eq:evolconstraint}
-\hbar^2\partial_\phi^2\Psi = \big(\hat{\mathcal H}_0^2+\hat\Theta_2\big)\Psi,
\end{equation}
where
\begin{equation}
\hat{\mathcal H}_0^2 = \frac{3}{4\pi G\gamma^2}\hat\Omega_0^2-m^2\hat V^2\phi^2,
\end{equation}
and $\hat\Theta_2$ is essentially $-(8\pi G)^{-1}\left(\sum_{\vec n,\epsilon}\hat{\mathcal C}_2^{\vec n,\epsilon}\right)^\dagger$, replacing the operator $\hat\pi_\phi^2$ with $\hat{\mathcal H}_0^2$, according to Eq.~\eqref{eq:evolconstraint} up to the considered perturbative order. We will call $\hat{\mathcal H}_0$ the square root of $\hat{\mathcal H}_0^2$. In the particular case when the field is massless, $\hat{\mathcal H}_0$ is well defined for all values of $\phi$, and coincides up to a constant multiplicative factor with $|\hat\Omega_0|$. More generally, one might replace $\hat{\mathcal H}_0^2$ with its positive part, starting from the realization that no solution to Eq. \eqref{eq:evolconstraint} exists in the unperturbed system in the sector where the considered operator is negative. Returning to the massless case, the solutions of
the equations	
\begin{equation}\label{eq:pnfrequency}
-i\hbar\partial_\phi\chi_0 = \pm\hat{\mathcal H}_0\chi_0
\end{equation}
satisfy the unperturbed equation
\begin{equation}\label{eq:unperturbed}
-\hbar^2\partial_\phi^2\chi_0=\hat{\mathcal H}_0^2\chi_0.
\end{equation}
We assume that a similar treatment can be reproduced without basic obstructions when $m\neq0$. Note, however, that in the massive case the solutions to Eq.~\eqref{eq:pnfrequency} do not satisfy Eq.~\eqref{eq:unperturbed}, owing to the dependence of $\hat{\mathcal H}_0$ on $\phi$. Nevertheless, we can use the former Schr\"odinger-like equation to introduce a kind of interaction picture that simplifies the description of the quantum evolution. In the following, we restrict to positive-frequency ``solutions'', which correspond to choosing the positive sign in Eq.~\eqref{eq:pnfrequency}. Note that, for a self-adjoint $\hat{\mathcal H}_0$, negative-frequency solutions remain completely characterized and can be obtained by complex conjugation.

Thus, let us introduce the operator corresponding to the time-ordered exponential of $\hat{\mathcal H}_0$, namely
\begin{equation}
\hat U = \mathcal P\exp\left(\frac i\hbar\int_{\phi_0}^\phi\!\!d\phi'\hat{\mathcal H}_0(\phi')\right),
\end{equation}
with the symbol $\mathcal P$ denoting the time ordering. We assume that it is unitary and use it to change to an interaction picture in the usual way~\cite{Galindo}: $\Psi_{\mathrm I}=\hat U^\dagger\Psi$. Equation~\eqref{eq:evolconstraint} then reads
\begin{equation}\label{eq:interaction}
-\hbar^2\partial_\phi^2\Psi_{\mathrm I} = \big(\hat\Theta_{2,\mathrm I}+ i\hbar\partial_\phi\hat{\mathcal H}_{0,\mathrm I}\big)\Psi_{\mathrm I}+2i\hbar\hat{\mathcal H}_{0,\mathrm I}\partial_\phi\Psi_{\mathrm I},
\end{equation}
where the subindex $\mathrm I$ stands for the operator representation in the discussed interaction picture. For instance, we have $\hat\Theta_{2,\mathrm I}=\hat U^\dagger \hat\Theta_{2} \hat U$. In addition, it is important to note that $\hat\Theta_2$ contains first-order derivatives with respect to $\phi$. We make this explicit by writing $\hat\Theta_2={}^{(0)}\hat\Theta_2-i\hbar{}^{(1)}\hat\Theta_2\partial_\phi$, where neither $^{(0)}\hat\Theta_2$ nor $^{(1)}\hat\Theta_2$ include differentiation with respect to $\phi$. If $m$ vanishes, so does $^{(1)}\hat\Theta_2$, simplifying the treatment considerably.

We can extract more information of the above equation by adopting a kind of Born-Oppenheimer approximation. Firstly, we assume that the solutions have the factorized form $\Psi(\phi,v,\mathcal N)=\chi_0(\phi,v)\psi(\phi,\mathcal N)$, where $\chi_0$ is a (positive-frequency) solution of Eq.~\eqref{eq:pnfrequency}. Inserting this ansatz in Eq.~\eqref{eq:interaction}, and taking the background inner product with $\chi_{0,\mathrm I}$, we arrive at
\begin{equation}\label{eq:ansatzeq}
-\hbar^2\partial_\phi^2\psi = \langle{}^{(0)}\hat\Theta_2+{}^{(1)}\hat\Theta_2\hat{\mathcal H}_0\rangle_{\chi_0}\psi+i\hbar\langle\partial_\phi\hat{\mathcal H}_0\rangle_{\chi_0}\psi+2i\hbar\langle\hat{\mathcal H}_0\rangle_{\chi_0}\partial_\phi\psi-i\hbar\langle{}^{(1)}\hat\Theta_2\rangle_{\chi_0}\partial_\phi\psi,
\end{equation}
where we have introduced the notation
\begin{equation}\label{eq:inhomwf}
\langle\hat{\mathcal O}(\phi)\rangle_{\chi_0} = \frac{\sum_v\chi_0^*(\phi,v)\hat{\mathcal O}(\phi)\chi_0(\phi,v)}{\sum_v|\chi_0(\phi,v)|^2} = \langle\hat{\mathcal O}_{\mathrm I}(\phi)\rangle_{\chi_{0,\mathrm I}},
\end{equation}
$\hat{\mathcal O}$ being a generic operator. We now focus our discussion on Eq.~\eqref{eq:ansatzeq} by neglecting all those nondiagonal contributions in Eq.~\eqref{eq:interaction} that mix $\chi_0$ with other background states. In this way, we get a second-order differential equation in the time $\phi$ that dictates the quantum evolution of the inhomogeneities, described by the wavefunction $\psi$.

If the characteristic time scale of the homogeneous sector is much smaller than that of the inhomogeneous one, the term $-\hbar^2\partial_\phi^2\psi$ ought to be negligible in Eq.~\eqref{eq:ansatzeq}. Besides, $-i\hbar\langle{}^{(1)}\Theta_2\rangle_{\chi_0}\partial_\phi\psi$ should also be negligible compared to $2i \hbar \langle\hat{\mathcal H}_0\rangle_{\chi_0}\partial_\phi\psi$, because of the perturbative nature of the inhomogeneities. Then, defining $\tilde\psi=\langle\hat{\mathcal H}_0\rangle_{\chi_0}\psi$, Eq.~\eqref{eq:ansatzeq} transforms into
\begin{equation}\label{eq:evolutioneq}
-i\hbar\partial_\phi\tilde\psi = \frac12\frac{\langle{}^{(0)}\hat\Theta_2+{}^{(1)}\hat\Theta_2\hat{\mathcal H}_0\rangle_{\chi_0}}{\langle\hat{\mathcal H}_0\rangle_{\chi_0}}\tilde\psi.
\end{equation}
Naturally, the validity of this approximation should be checked once the solution is obtained. Within its regime of applicability, the evolution of the wavefunction of the inhomogeneities in the relational time $\phi$ is given by the first-order equation \eqref{eq:evolutioneq}. This expression is the analogue of Eq.~(4.12) of Ref.~\cite{AAN2} (see also the extension to the massive case in Ref.~\cite{AAN3}). Note, however, that the function $S_2^{\prime(\mathcal Q)}$ (from which the operator analogous to $\hat\Theta_2$ is derived in those references) differs from our second-order constraint because we have used a different parametrization of the inhomogeneous sector---selected by the criteria of symmetry and \emph{unitary field dynamics}---and we have incorporated backreaction effects. Besides, the effective Hamiltonian that one gets from Eq.~\eqref{eq:evolutioneq} is---in the massless case---$\frac12\langle\hat H_0\rangle_{\chi_0}^{-1}\langle\hat\Theta_2\rangle_{\chi_0}$, instead of $\tfrac12\langle\hat H_0^{-1/2}\hat\Theta_2\hat H_0^{-1/2}\rangle_{\chi_0}$.\footnote{In addition, notice the possible difference in the expectation value on the background coming from the specific prescription employed in the loop quantization of the FLRW geometry.}

For background states $\chi_0$ that are highly peaked about a dynamical, effective trajectory, the result of taking the expectation value in the above Hamiltonian amounts to the evaluation of the homogeneous variables in the corresponding effective background. We then reach a description for the inhomogeneities that reproduces a quantum field theory in an effective curved spacetime. In addition, note that, since the corresponding Hamiltonian for the inhomogeneities is time (i.e., $\phi$) dependent, the vacuum state that we have selected is not conserved under such an evolution. These facts can well produce observable effects in the cosmological scalar perturbations, that we plan to investigate in a future work~\cite{FMOfu}.

\subsection{Recursive characterization}\label{subsec:recursive}

Even if the field could not be used as a relational time, the fact that the quantum constraint is a difference operator in $v$ and that the volume operator is bounded from below in each superselection sector, with a definite triad orientation, allow us to characterize the quantum solutions in each of these sectors $\mathcal H_\mathrm{\varepsilon}^\pm$ by their ``initial data'' on the minimum-volume section $|v|=\varepsilon$. We assume here a perturbative expansion for the physical states and truncate all the expressions at second order. The procedure is entirely analogous to that of Ref.~\cite{FMO12}; we sketch it here for completeness.

Up to higher-order terms, we can expand perturbatively the wavefunction of a quantum state as
\begin{equation}
\Psi(\phi,v,\mathcal N) = \Psi^{(0)}(\phi,v,\mathcal N)+\Psi^{(2)}(\phi,v,\mathcal N).
\end{equation}
Now, we can expand the constraint equation $(\Psi| \hat {\mathcal C}^{\dagger}=0$ order by order. The zeroth-order piece tells us that $(\Psi^{(0)}|$ is a solution to the unperturbed Hamiltonian constraint, i.e., it satisfies the second-order difference equation $(\Psi^{(0)}|\hat {\mathcal C}_0=0$. Owing to the properties of $\hat\Omega_0^2$, this means that the value of the wavefunction $\Psi^{(0)}$ at every $v\in\mathcal L_{\varepsilon}^\pm$ can be determined given the initial value on the corresponding minimum-volume section, $\Psi^{(0)}(\phi,\pm\varepsilon,\mathcal N)$. In this way, all the solutions in, e.g., the superselection sector $\mathcal H_\varepsilon^+$ are characterized by their initial data at $v=\varepsilon$.

The second-order piece of the constraint equation is
\begin{equation}
\big(\Psi^{(2)} \big |\hat {\mathcal C}_0+\big(\Psi^{(0)}\big | \Big(\sum_{\vec n,\epsilon}\hat {\mathcal C}^{\vec n,\epsilon}_2\Big)^\dagger = 0.
\end{equation}
Therefore, $\Psi^{(2)}$ satisfies a second-order difference equation as $\Psi^{(0)}$, but with a source term that is sensitive to the information about the inhomogeneities contained in the zeroth-order state $\Psi^{(0)}$. Again, since $\hat\Omega_0^2$ and $\hat\Lambda_0$ decouple the superselection sectors $\mathcal H_\varepsilon^\pm$, the knowledge of $\Psi^{(2)}$ on the section $|v|=\varepsilon$ suffices to determine its value in the rest of the semilattice $\mathcal L_\varepsilon^\pm$. Hence, we can identify the solutions to the constraint with their initial data. The vector space of these data can be endowed with an inner product by applying the so-called reality conditions~\cite{Rendall} to a complete set of observables. In this way, one obtains a physical Hilbert space which is equivalent to $\mathcal H_\mathrm{kin}^\mathrm{matt}\otimes\mathcal F$.

\section{Summary and conclusions}\label{sec:conclusions}

We have achieved a complete quantization of an FLRW universe provided with a massive scalar field with scalar perturbations, in the case in which the spatial slices are compact and flat. The strategy has been the same as in the case of a universe with spatial sections of positive curvature~\cite{FMO12}. We have truncated the action at quadratic order in the perturbations, fixed the gauge at the classical level (by two different sets of conditions), and performed a canonical transformation, scaling the field by the background scale factor and changing its conjugate momentum, following the uniqueness results of Refs.~\cite{CMOV11,CMOV12, CCMMV12, FMOV12} in order to reach a privileged description of the system. Then, we have constructed the kinematical Hilbert space by combining a preferred Fock representation of the local degrees of freedom with the LQC representation of the homogeneous background. We expect this hybrid approach to be a valid approximation as long as the effects of quantum geometry become significant only in the homogeneous sector.

In the kinematical Hilbert space, we have promoted the Hamiltonian constraint to an operator, following a prescription inspired by the unperturbed case. We have ensured that the superselection sectors of the unperturbed theory remain superselected. The constraint equation that is to be satisfied by the physical states can be interpreted as a second-order evolution equation in the relational time provided by the zero mode of the scalar field. If one admits a kind of Born-Oppenheimer approximation (which assumes that the background and the inhomogeneous sector have very different characteristic ``time'' scales), this can be translated into a first-order evolution equation for the wavefunction of the inhomogeneities. Alternatively, quantum solutions to the constraint equation can be characterized in each superselection sector in terms of their initial data on the minimum-volume section. The recurrence relation that fixes their value on the remaining volume sections becomes especially simple if a perturbative expansion for them is supposed. The physical Hilbert space can then be built by using reality conditions in order to endow the space of initial data with a suitable inner product. As in the Gowdy models~\cite{hybrid1}, the Fock space of the inhomogeneities is recovered as one of the factors in the tensor product that gives this physical Hilbert space.

It is worth emphasizing that, compared to other approaches in the literature~\cite{AAN1,AAN2,AAN3} that incorporate features common to the hybrid quantization, our approach succeeds in providing (i) a symplectic structure and a consistent description in terms of constraints for the perturbed FLRW system, both before and after gauge fixing and reduction, (ii) a complete quantization independently of the use of the homogeneous scalar field as a relational time, (iii) a second-order evolution equation for the inhomogeneities when such a relational time is adopted, rather than a first-order one, and (iv) a {\emph{unitary}} quantum field theory of the cosmological perturbations in a dressed background, in appropriate regimes.

Let us also remark that, even tough we have fixed the gauge in the classical theory, we have reached some reassuring results in our discussion, pointing to the gauge independence of the quantization. In the gauge of flat spatial sections, the scaled perturbation of the matter field coincides with a gauge-invariant quantity---the widely known MS variable~\cite{Mukhanov05}. Although that is not true in the longitudinal gauge, in that case the quantization performed in terms of the scaled field perturbation (and an appropriate conjugate momentum) is unitarily equivalent to the one which takes as the fundamental canonical pair the MS variable and its time derivative. Yet, it may be asked what we have gained with our choice of field variables, since the MS variable is probably the most used one in the literature on inflation in flat FLRW models, owing to its diagonal Lagrangian, gauge invariance, and good ultraviolet behavior. Note, however, that our way to pick out the chosen representation has been completely different. We have required a quantization with a translation-invariant vacuum state and unitarily implementable field dynamics in the (classical) homogeneous background, and these criteria have led us to the preferred variables and a \emph{unique} unitary equivalence class of Fock representations for them. Any nontrivial mode dependent (and hence \emph{nonlocal}), time-varying linear canonical transformation of the field variables would lead to a field description in which, a priori, there exists no well established reason to demand those criteria.

In the process of reduction of the classical system and in the definition of the new variables~\eqref{eqs:newvariablesA} and \eqref{eqs:newvariablesB} (in the corresponding gauges), we have needed to correct the homogeneous variables with second-order terms. Without these corrections, which can be interpreted as a backreaction, we would not have been able to perform successfully neither the partial deparametrization in the longitudinal gauge nor the transformation that includes the scaling of the matter field perturbation, which has permitted us to attain the privileged quantization. Although these terms are subdominant in the perturbative regime compared with the homogeneous ones, they introduce corrections in the dynamics.

The model is now ready to be analyzed and simulated numerically so as to obtain predictions, e.g., for the power spectrum of primordial perturbations that can be compared with the available observational data~\cite{wmaplanck}. In this way one would be able to check whether the quantization put forward is physically acceptable and, moreover, face the challenge of seeking for departures from the standard results in cosmology which could be attributed to quantum geometry effects and that, in spite of being small, might nonetheless be falsified within the constringent margins of observational error. These and other issues will be the object of future investigation.

\section*{Acknowledgements}

We would like to thank G. Calcagni, D. Mart\'\i{}n-de Blas, and M. Mart\'{\i}n-Benito for discussions. This work was supported by the Project No.\ MICINN/MINECO FIS2011-30145-C03-02 from Spain. M.F.-M. acknowledges CSIC and the European Social Fund for support under the grant JAEPre\_2010\_01544. J.O. acknowledges Pedeciba.

\appendix

\section{Unreduced Hamiltonian}\label{sec:Hamiltonian}

The structure of the Hamiltonian, truncated at quadratic order in the perturbations, is given by Eq.~\eqref{eq:Hamiltonian}. The expression of the zeroth-order Hamiltonian $H_{|0}$ can be found in Eq.~\eqref{eq:H_0}, while the higher-order terms are
\begin{subequations}
\begin{align}
H^{\vec n,\epsilon}_{|2} &= \frac12e^{-3\alpha}\Big\{ -\pi_{a_{\vec n,\epsilon}}^2+\pi_{b_{\vec n,\epsilon}}^2+\pi_{f_{\vec n,\epsilon}}^2+2\pi_\alpha(a_{\vec n,\epsilon}\pi_{a_{\vec n,\epsilon}}+4b_{\vec n,\epsilon}\pi_{b_{\vec n,\epsilon}})-6\pi_\varphi a_{\vec n,\epsilon}\pi_{f_{\vec n,\epsilon}} \nonumber\\
&\phantom{=\frac12e^{-3\alpha}\Big\{} +\pi_\alpha^2\Big(\tfrac12a_{\vec n,\epsilon}^2+10b_{\vec n,\epsilon}^2\Big)+\pi_\varphi^2\Big(\tfrac{15}2a_{\vec n,\epsilon}^2+6b_{\vec n,\epsilon}^2\Big) \nonumber\\
&\phantom{=\frac12e^{-3\alpha}\Big\{} -e^{4\alpha}\Big[\tfrac13\omega_n^2a_{\vec n,\epsilon}^2+\tfrac13(\omega_n^2-18)b_{\vec n,\epsilon}^2+\tfrac23\omega_n^2a_{\vec n,\epsilon}b_{\vec n,\epsilon}-\omega_n^2f_{\vec n,\epsilon}^2\Big] \nonumber\\
&\phantom{=\frac12e^{-3\alpha}\Big\{} +e^{6\alpha}\tilde m^2\Big[3\varphi^2\Big(\tfrac12a_{\vec n,\epsilon}^2+2b_{\vec n,\epsilon}^2\Big)+6\varphi a_{\vec n,\epsilon}f_{\vec n,\epsilon}+f_{\vec n,\epsilon}^2\Big]\Big\}, \\
H^{\vec n,\epsilon}_{|1} &= \frac12e^{-3\alpha}\Big[-2\pi_\alpha\pi_{a_{\vec n,\epsilon}}+2\pi_\varphi\pi_{f_{\vec n,\epsilon}}-\big(\pi_\alpha^2+3\pi_\varphi^2\big)a_{\vec n,\epsilon}-\tfrac23\omega_n^2e^{4\alpha}(a_{\vec n,\epsilon}+b_{\vec n,\epsilon}) \nonumber\\
&\phantom{= \frac12e^{-3\alpha}\Big[} +e^{6\alpha}\tilde m^2\varphi(3\varphi a_{\vec n,\epsilon}+2f_{\vec n,\epsilon})\Big], \label{eq:H^n_|1}\\
H^{\vec n,\epsilon}_{\_1} &= \frac13e^{-\alpha}\big[-\pi_{a_{\vec n,\epsilon}}+\pi_{b_{\vec n,\epsilon}}+\pi_\alpha(a_{\vec n,\epsilon}+4b_{\vec n,\epsilon})+3\pi_\varphi f_{\vec n,\epsilon}\big]. \label{eq:H^n__1}
\end{align}
\end{subequations}
The vanishing of the spatial curvature in the FLRW sector changes some coefficients with respect to the case of spatial sections with positive curvature, considered in Refs.~\cite{HH85,FMO12}. Only the expression of $H^{\vec n,\epsilon}_{\_1}$ is formally the same, although it is included here for the sake of completeness.

\section{Another gauge-invariant canonical pair}\label{sec:gauge}

In an approximately flat universe filled with a scalar field, the MS variable satisfies a KG equation with time-dependent mass. This fact makes it especially convenient for the quantum treatment of the inhomogeneities during inflation. However, this is no longer the case when the spatial sections of the background geometry are not flat. This explains why we did not consider a quantization in terms of the MS variable in previous works~\cite{FMOV12,FMO12}. Rather, we considered a particular combination of the gauge-invariant energy-density and matter-velocity perturbations defined by Bardeen~\cite{Bardeen80}. These are given, respectively, by
\begin{subequations}\label{eqs:invariant}
\begin{align}
\mathcal E^m_{\bar n} &= \frac{e^{-2\alpha}}{E_0}\left[-\dot\varphi^2g_{\bar n}+\dot\varphi\dot f_{\bar n}+(e^{2\alpha}m^2\varphi+3\dot\alpha\dot\varphi)f_{\bar n}\right], \\
v^s_{\bar n} &= \frac{1}{\omega_{n}}\left[\frac{\omega_{n}^2}{\dot\varphi}f_{\bar n}+\left(\frac{k_{\bar n}}{N_0}-3\dot b_{\bar n}\right)\right],
\end{align}
\end{subequations}
where $E_0=e^{-2\alpha}(\dot\varphi^2+e^{2\alpha}\tilde m^2\varphi)/2$ and ${\bar n}$ is a label in a real eigenbasis $\{{\tilde Q}_{\bar n}\}$ of the LB operator, similar to the label $({\vec n},\epsilon)$ of the flat case. The above expressions hold formally regardless of the curvature of the spatial sections if the ADM variables are expanded in the eigenbasis $\{{\tilde Q}_{\bar n}\}$, in an analogous way to what is done in Eqs.~\eqref{eqs:expansions}. The eigenvalue corresponding to the mode ${\tilde Q}_{\bar n}$ is denoted by $-\omega_{n}^2$ (again, different modes can have the same eigenvalue). The exact value of $\omega_n$ depends on the kind of spatial sections that the studied FLRW model possesses.

In terms of the phase space variables,
\begin{subequations}
\begin{align}
\mathcal E^m_{\bar n} &= \frac{e^{-6\alpha}}{E_0}\big[\pi_\varphi\pi_{f_{\bar n}}+(e^{6\alpha}\tilde m^2\varphi-3\pi_\alpha\pi_\varphi)f_{\bar n}-3\pi_\varphi^2a_{\bar n}\big], \\
v^s_{\bar n} &= \frac{1}{\omega_{n}}\left[\omega_{n}^2\frac{e^{2\alpha}}{\pi_\varphi}f_{\bar n}-3e^{-2\alpha}\left(\frac{\omega_{n}^2}{\omega_{n}^2-3k}\pi_{b_{\bar n}}+4\pi_\alpha b_{\bar n}\right)\right],
\end{align}
\end{subequations}
where $k=0,\pm1$ is the curvature parameter of the background FLRW model. Naturally, we can obtain new gauge-invariant quantities from these ones. In particular, we are interested in the combinations
\begin{subequations}\label{eqs:Psi_n1}
\begin{align}
\Psi_{\bar n} &= \frac{1}{\omega_{n}}\frac{e^{5\alpha}}{\pi_\varphi}E_0\mathcal E^m_{\bar n}, \\
\Pi_{\Psi_{\bar n}} = \dot\Psi_{\bar n} &= -\frac{\sqrt{\omega_{n}^2-3k}}{\omega_{n}}e^{-\alpha}\pi_\varphi v^s_{\bar n}+\frac{e^{-2\alpha}}{\pi_\varphi}\big(e^{6\alpha}m^2\varphi-2\pi_\alpha\pi_\varphi\big)\Psi_{\bar n}.
\end{align}
\end{subequations}
Remarkably, the variables $\Psi_{\bar n}$ satisfy the equations for the modes of a KG field with time-dependent mass, irrespective of the curvature of the spatial sections.

In both of the gauge fixings considered here, the expressions of $\Psi_{\bar n}$ and its derivative reduce to the form
\begin{subequations}
\begin{align}
\Psi_{\bar n} &= \frac{1}{\sqrt{\omega_{n}^2-3k}}\left(\bar\pi_{\bar f_{\bar n}}+\chi\bar f_{\bar n}\right), \\
\Pi_{\Psi_{\bar n}} &= \frac\chi{\sqrt{\omega_{n}^2-3k}}(\bar\pi_{\bar f_{\bar n}}+\chi\bar f_{\bar n})-\sqrt{\omega_{n}^2-3k}\bar f_{\bar n},
\end{align}
\end{subequations}
up to the considered perturbative order. The variable $\chi$ is a function of the homogeneous variables whose explicit expression depends on the gauge. In the gauge of flat spatial sections, it is given by
\begin{equation}
\chi = \frac{e^{-2\bar\alpha}}{\pi_{\bar \varphi}}\left(3\frac{\pi_{\bar \varphi}^3}{\pi_{\bar\alpha}}+e^{6\bar\alpha}\tilde m^2 \bar \varphi-2\pi_{\bar\alpha}\pi_{\bar \varphi}\right),
\end{equation}
while in the longitudinal gauge it adopts the simpler form
\begin{equation}
\chi = \frac{e^{-2\bar\alpha}}{\pi_{\bar \varphi}}\left(e^{6\bar\alpha}\tilde m^2 \bar \varphi-2\pi_{\bar\alpha}\pi_{\bar \varphi}\right).
\end{equation}
Note that the definition of $\pi_{\bar f_{\bar n}}$ is also different in the two cases. From these equations, it is clear that $\Psi_{\bar n}$ and $\Pi_{\Psi_{\bar n}}$ are canonically conjugate in the reduced phase spaces.

For this parametrization of the field, there is a preferred Fock quantization selected by the criteria of (spatial) symmetry invariance of the vacuum and unitary implementability of the dynamics. In the closed ($k=+1$) case, it was shown that the quantization corresponding to these variables is unitarily equivalent to the one constructed for the scaled perturbation of the matter field by applying the uniqueness criteria~\cite{FMOV12, FMO12}.

The same result is obtained in the flat ($k=0$) case. Let us adopt the annihilation-like variables that would be natural if the field were massless,
\begin{equation}
a_{\Psi_{{\vec n},\epsilon}} = \frac1{\sqrt{2\omega_n}}(\omega_n\Psi_{{\vec n},\epsilon}+i\Pi_{\Psi_{{\vec n},\epsilon}}).
\end{equation}
Here, we have made explicit the identification of the label ${\bar n}$ with $({\vec n},\epsilon)$ in the flat case.
Of course, there is a symplectomorphism relating these variables with the ones corresponding to the modes $\bar f_{\vec n, \epsilon}$ and their momenta:
\begin{equation}
a_{\Psi_{{\vec n},\epsilon}} = \tilde\lambda_n^+a_{\bar f_{{\vec n},\epsilon}}+\tilde\lambda_n^-a_{\bar f_{{\vec n},\epsilon}}^*.
\end{equation}
Once again, the unitary implementability of this transformation relies on the asymptotic behavior of the coefficients $\tilde\lambda_n^-$. It is easy to check that $\tilde\lambda_ n^-=i\chi^2/(2\omega_n^2)$, and hence the sequence $\{G_n |\tilde\lambda_n^-|^2\}$ is summable, and the alternate quantizations are unitarily related.

\end{document}